\documentstyle[12pt]{article}
\newcommand {\cD}{{\cal D}}

\newcommand {\cK}{{\cal K}}

\newcommand {\cP}{{\cal P}}

\newcommand {\cV}{{\cal V}}

\def\a{\alpha}
\def\b{\beta}

\def\d{\delta}

\def\g{\gamma}
\def\G{\Gamma}

\def\l{\lambda}
\def\m{\mu}

\def\q{\theta}

\newcommand{\ad}{{\dot{\alpha}}}                           
\newcommand{\bd}{{\dot{\beta}}}                            
\newcommand{\ve}{\varepsilon}                            
\newcommand{\vb}{{\bar\varepsilon}}                            
\newcommand{\mb}{{\bar\mu}}                            

\renewcommand{\aa}{{\a\ad}}

\newcommand{\kl}{{\a(k)\ad(l)}}

\renewcommand{\ss}{{\a(s)\ad(s)}}
\newcommand{\rr}{{\a(s-1)\ad(s-1)}}

\newcommand{\sr}{{\a(s)\ad(s-1)}}
\newcommand{\rs}{{\a(s-1)\ad(s)}}
\newcommand{\ro}{{\a(s-1)\ad(s-2)}}
\newcommand{\oo}{{\a(s-2)\ad(s-2)}}
\newcommand{\os}{{\a(s-2)\ad(s)}}

\newcommand{\cDB}{{\bar{\cal D}}}
\newcommand{\pa}{\partial}                           
\newcommand{\hf}{\frac12}

\newcommand{\be}{\begin{equation}}
\newcommand{\ee}{\end{equation}}
\newcommand{\bea}{\begin{eqnarray}}
\newcommand{\eea}{\end{eqnarray}}
\newcommand{\non}{\nonumber}

\begin{document}
\begin{titlepage}

\begin{flushright}
UMDEPP 97-23 \\
TSU-QFT-13/96 \\
\end{flushright}

\begin{center}
\large{{\bf N = 2 Supersymmetry of Higher Superspin Massless
Theories} } \\
\vspace{1.0cm}

\large{ S. James Gates Jr.\footnote{ E-mail: gates@umdhep.umd.edu.}  } \\

\footnotesize{{\it Department of Physics, University of Maryland \\
College Park, MD 20742-4111, USA}} \\
\vspace{0.5cm}

\large{Sergei M. Kuzenko\footnote{ E-mail: kuzenko@phys.tsu.tomsk.su.}
and Alexander G. Sibiryakov\footnote{ E-mail: sib@phys.tsu.tomsk.su.}
 } \\

\footnotesize{{\it Department of Quantum Field Theory, Tomsk State
University \\
Lenin Ave. 36, Tomsk 634050, Russia} \\
 }
\end{center}
\vspace{1.5cm}

\begin{abstract}
We present $N=2$ supersymmetry transformations, both in $N=1$, $D=4$
Minkowski and anti-de Sitter superspaces, for higher superspin massless
theories. It is noted that the existence of dual versions of massless
supermultiplets with arbitrary superspin $s$ may provide a basis for
understanding duality in $N=1$, $D=4$ superstring theory. We further
conjecture that the $N=1$, $D=4$ supergravity pre-potential together with
all higher superspin $s$ pre-potentials are components of an
$N=1$, $D=4$ superstring pre-potential.
\end{abstract}
\vspace{10mm}

\begin{flushleft}
September 1996
\end{flushleft}

\vfill
\null
\end{titlepage}

\newpage

\noindent
{\bf 1.} Recently, the linearized Lagrangian gauge theory formulations 
for arbitrary higher superspin massless multiplets have been constructed
in $N=1$, $D=4$ Minkowski superspace \cite{kps,ks1} (see also \cite{bk})
and then extended to the case of anti-de Sitter (AdS) supersymmetry 
\cite{ks2}. These models are realized in terms of real unconstrained 
superfield pre-potentials (being the higher spin analogues of the
gravitational axial vector superfield \cite{os}) along with complex 
constrained ones: longitudinal linear superfields or transversal linear 
superfields (non-trivial analogues of the chiral compensator in the 
old minimal supergravity \cite{sg} and the complex linear compensator 
in the nonminimal supergravity \cite{gs}, respectively). Transversal 
and longitudinal linear superfields are known to be native inhabitants 
of the AdS superspace \cite{is}.  Combining massless multiplets of all 
superspins in the AdS superspace, one probably results in a theory 
\cite{ks3} possessing an infinite-dimensional symmetry algebra whose
finite-dimensional subalgebra includes the $N=2$ AdS superalgebra 
$osp(4,2)$. In this respect, it would be of interest to analyze possible 
realizations of $N=2$ supersymmetry for higher superspin massless 
multiplets.

A major motivation to pursue the approach of references \cite{kps,ks1} 
is to develop a new tool with which to analyze superstring theory.  It 
should be clearly stated that the theories developed thus far 
\cite{kps,ks1,ks2} are {\it {not}} linearized $D=4$ superstrings but 
are a more general class of higher superspin theories. There is no 
impediment to adding together arbitrary numbers of these linearized 
actions with arbitrary values of $s$ and taking the $s \to \infty$ 
limit.  However, with the proper choice of multiplicities and spectrum 
$N=1$, $D=4$, superstrings should emerge as special cases.

{\bf 2.}
The formulations of Refs. \cite{kps,ks1} in $N=1$, $D=4$ Minkowski
superspace (with standard covariant derivatives $D_A = (\pa_a,
D_\a,\bar D^\ad)$) involve so-called transversal and longitudinal 
linear superfields. A complex tensor superfield $\G(k,l)$ subject
to the constraint\footnote{Throughout the paper we consider only 
Lorentz tensors symmetric in their undotted indices and separately 
in their dotted ones.   For a tensor of type $(k,l)$ with $k$ undotted 
and $l$ dotted indices we use the shorthand notations
$
\Psi(k,l) \equiv \Psi_\kl   \equiv \Psi_{\a_1 \ldots
\a_k\ad_1\ldots \ad_l} = \Psi_{(\a_1 \ldots \a_k)(\ad_1\ldots
\ad_l)}$.
Following Ref. \cite{vas} we assume that the indices,
which are denoted by one and the same letter,
should be symmetrized separately with respect to upper and
lower indices; after the symmetrization, the maximal possible number
of the upper and lower indices denoted by the same letter
are to be contracted. In particular
$
\phi_{\a(k)} \psi_{\a(l)}
\equiv \phi_{(\a_1\ldots\a_k} \psi_{\a_{k+1}\ldots\a_{k+l})}$,
$
\xi^\a \phi_{\a(k)}
\equiv  \xi^\b \phi_{(\b\a_1\ldots\a_{k-1})}$.
Given two tensors of the same type, their contraction is denoted by
$f \cdot g \equiv f^\kl g_\kl$.
}
\bea
\bar D^\ad \G_\kl & = & 0 ~~~, \qquad l>0 ~~~, \non \\
\bar D^2 \G_{\a (k)} & = & 0 ~~~,\qquad l=0 ~~~,
\label{3}
\eea
is said to be transversal linear. A longitudinal linear superfield
$G(k,l)$  is defined to satisfy the constraint
\be
\bar D_\ad G_\kl=0 ~~~,
\label{4}
\ee
(the symmetrization over all dotted indices is assumed).
The above constraints imply that
$\G(k,l)$ and $G(k,l)$ are linear in the usual sense
\be
\bar D^2 \G(k,l) = \bar D^2 G(k,l) = 0 ~~~.
\label{5}
\ee

The constraints (\ref{3}) and (\ref{4}) can be resolved in terms of
unconstrained superfields
\bea
 \G_\kl &=& \bar D^\ad \Phi_{\a(k)\ad(l+1)} ~~~,
\label{6}                                  \\
 G_\kl &=& \bar D_\ad \Psi_{\a(k)\ad(l-1)} ~~~.
\label{7}
\eea
Here the superfields $\Phi$ and $\Psi$ are defined modulo arbitrary shifts
of the form
\bea
 \d \Phi(k,l+1) &=& \g(k,l+1) ~~~,
\label{8}\\
 \d \Psi(k,l-1) &=& g(k,l-1) ~~~,
\label{9}
\eea
involving a transversal linear superfield $\g(k,l+1)$ and a longitudinal 
linear superfield $g(k,l-1)$.  It follows that any transversal linear 
superfield $\G(k,l)$, which appears in a supersymmetric field theory, 
can be equivalently replaced by an unconstrained one, via rule (\ref{6}),
that introduces the additional gauge invariance (\ref{8}). The gauge 
parameter, which is a transversal linear superfield, can also be re-expressed
in terms of an unconstrained superfield introduced via the same rule,
and on and on. The number of dotted indices on the superfields increases
indefinitely in this process, thus providing us with a gauge structure of 
infinite stage reducibility. Any longitudinal linear superfield $g(k,l)$
can be replaced by an unconstrained one in correspondence with the rule 
(\ref{7}). Similarly, the longitudinal linear gauge parameter in Eq. 
(\ref{9}) can be re-expressed in terms of an unconstrained one and so on. 
This results in a gauge structure of  $l-1$ stage reducibility.

Two formulations for the massless multiplet of a half-integer superspin 
$s+1/2$ ($s=1,2,\ldots$) which were called in Ref. \cite{kps} transversal
and longitudinal, contain the following dynamical variables respectively:
\bea
\cV^\bot_{s+1/2}& = &\left\{H(s,s), \quad \G(s-1,s-1),
\quad \bar{\G}(s-1,s-1)\right\} ~~~,    \\
\label{10}
\cV^\|_{s+1/2} &=& \left\{H(s,s), \quad G(s-1,s-1),
\quad \bar{G}(s-1,s-1)\right\} ~~~.
\label{11}
\eea
Here $H(s,s)$ is real, $\G(s-1,s-1)$ transversal linear and $G(s-1,s-1)$ 
longitudinal linear tensor superfields.  The case $s=1$ corresponds to 
the supergravity multiplet investigated in detail in literature (see 
\cite{bk,ggrs} for a review).  Two formulations of Ref. \cite{ks1} for 
the massless multiplet of an integer superspin $s$, ($s=1,2,\ldots$)
longitudinal and transversal, contain the following dynamical variables 
respectively:
\bea
\cV^\bot_s &=& \left\{H'(s-1,s-1), \quad \G'(s,s),
\quad \bar{\G'}(s,s)\right\} ~~~,\\
\label{12}
\cV^\|_s &=& \left\{H'(s-1,s-1), \quad G'(s,s),
\quad \bar{G'}(s,s)\right\} ~~~.
\label{13}
\eea
Here $H'(s-1,s-1)$ is real, $\G'(s,s)$ transversal linear and
$G'(s,s)$ longitudinal linear tensor superfields. The case $s=1$
corresponds to the gravitino multiplet \cite{ks1,o77,fvdv,gs1}
(see \cite{bk,ggrs} for a review).

In the transversal half-integer-superspin formulation, the action functional 
reads
\bea
S^\bot_{s+1/2}&=&\left( -\hf \right)^s \displaystyle\int d^8z \Big\{
\frac{1}{8} H^\ss  D^\b \bar D^2 D_\b H_\ss \cr \non \\
&+& H^\ss \left( D_\a \bar D_\ad \G_\rr
-\bar D_{\ad}  D_\a \bar\G_\rr \right)\cr \non \\
&+&\left( \bar\G \cdot \G
+ \left(\frac{s+1}s\right) \G \cdot \G + {\rm c.c.}\right)\Big\} ~~~.
\label{14}
\eea
In the longitudinal formulation, the action takes the form
\bea
S^\|_{s+1/2}&=&\left(-\hf\right)^s \displaystyle\int d^8z \Big\{
\frac 18 H^\ss  D^\b \bar D^2
 D_\b H_\ss  \non \\
&-& \frac{1}{8} \left( \frac{s}{2s+1} \right) \Big[ \,
\big[ D_\a, \bar D_{\ad}\big] H^\ss \, \Big]  \Big[ \,
\big[ D^\b, \bar D^{\bd}\big]
H_{\b\a(s-1)\bd\ad(s-1)} \, \Big] \non \\
&+& \left(\frac{s}{2}\right) \Big[ \partial_{\a\ad} H^\ss \Big] \Big[
\partial^{\b\bd}
H_{\b\a(s-1)\bd\ad(s-1)} \Big]   \non \\
&+&  \left( \frac{2is}{2s+1}  \right)  \pa_{\a\ad} H^\ss
\Big( G_\rr - \bar G_\rr \Big)  \non \\
&+& \frac{1}{2s+1} \left( \bar G \cdot G - \frac{s+1}s G \cdot G
+ {\rm c.c.}\right)\Big\}   ~~~.
\label{15}
\eea
The gauge transformations for the superfields $H(s,s)$, $\G(s-1,s-1)$
and $G(s-1,s-1)$  are given in the form
\bea
\d _g H(s,s) &=& g(s,s) + \bar g(s,s) ~~~,
\label{16}\\
\d _g \G_\rr  &=& \frac{s}{2(s+1)} \bar D^\ad D^\a \bar g_\ss ~~~,
\label{17}\\
\d _g G_\rr  &=& \frac s{2(s+1)} D^\a \bar D^\ad g_\ss + is \pa^\aa g_\ss ~~~,
\label{18}
\eea
with a longitudinal linear parameter $g(s,s)$.
The action $S^\bot_{s+1/2}$ is invariant under the gauge transformations
(\ref{16}) and (\ref{17}), and $S^\|_{s+1/2}$ is invariant under
the gauge transformations
(\ref{16}) and (\ref{18}). The actions (\ref{14}) and (\ref{15}) are
dually equivalent \cite{kps}.

The massless multiplet of integer superspin $s$ is described in the
longitudinal formulation by the action functional
\bea
S^\|_s&=&\left(-\hf\right)^s \displaystyle\int d^8z \Big\{
\frac18 H'^\rr  D^\b \bar D^2 D_\b H'_\rr \non \\
&+& \left(\frac{s}{s+1}\right) H'^\rr \left( D^\a \bar D^\ad G'_\ss -
\bar D^{\ad} D^\a \bar G'_\ss \right) \non \\
&+&\left( \bar G' \cdot G'
+ \left(\frac{s+1}s\right) G' \cdot G'
+ {\rm c.c.}\right)\Big\} ~~~,
\label{19}
\eea
while the action in the transversal formulation takes the form
\bea
S^\bot_s&=&-\left(-\hf\right)^s \displaystyle\int d^8z \Big\{-
\frac{1}{8} H'^\rr  D^\b \bar D^2
 D_\b H'_{\a(s-1)\ad(s-1)}  \non \\
&+& \frac{1}{8} \left( \frac{s^2}{(s+1)(2s+1)} \right)
 \Big[ \, \big[ D^\a, \bar D^{\ad}\big] H'^\rr \Big] \, \Big[ \,
\big[ D_{\a}, \bar D_{\ad}\big] H'_\rr \, \Big] \non \\
&+& \hf \left( \frac{s^2}{s+1} \right) \Big[ \pa^\aa H'^\rr \Big]
\, \Big[ \pa_\aa H'_\rr \Big]  \non \\
&+& \left( \frac{2is}{2s+1} \right) H'^\rr
\pa^\aa \left(\G'_\ss - \bar\G'_\ss \right) \non \\
&+&\left( \frac{1}{2s+1} \right) \left( \bar\G' \cdot \G'
- \left( \frac{s+1}{s} \right) \G' \cdot \G' + {\rm c.c.}\right)\Big\}
~~~.
\label{20}
\eea
The gauge freedom for the superfields $H'(s-1,s-1)$, $G'(s,s)$
and $\G'(s,s)$ reads
\bea
 \d _\g H'(s-1,s-1) &=& \g(s-1,s-1) + \bar \g(s-1,s-1) ~~~,
\label{21}\\
\d _\g G'_\ss &=& \hf \bar D_\ad D_\a \bar \g_\rr ~~~,
\label{22}\\
 \d _\g \G'_\ss &=& \hf D_\a \bar D_\ad \g_\rr -is \pa_\aa \g_\rr ~~~,
\label{23}
\eea
with a transversal linear parameter $\g(s-1,s-1)$.  So $S^\|_s$ is invariant 
under the gauge transformations (\ref{21}) and (\ref{22}), and $S^\bot_s$ 
is invariant under the gauge transformations (\ref{21}) and (\ref{23}).
The actions (\ref{19}) and (\ref{20}) are dually equivalent \cite{ks1}.

The appearance of these dual equivalent descriptions is very noteworthy.
It extends a property of superspace supergravity theory. There it is
known that duality transformations on the superspace are related to
changes of the Weyl compensating multiplet \cite{gros}. In 1985 \cite{gn}, 
it was proposed that this same type of duality transformation must exist
for dual descriptions of the $D=10$ heterotic string low-energy effective
action\footnote{This proposal was the first suggestion of string-string
duality in the literature.} and it was conjectured that this was likely
the case for the entire theory.  Presently duality is one of the most
intensively studied topics in superstring theory having led to many
new insights into the non-perturbative aspects of the theory.  Although
the original conjectures were for $D=10$ theories, the fact that these dual
actions described above exist for arbitrary values of $s$ provides additional
support for the now widely accepted belief in the presence of such dual
versions, at least for $N = 1$, $D=4$ superstrings.

The first lines in (\ref{16}) and (\ref{21}) together with the existence of 
the duality transformations discussed above lead to a simple interpretation.
The pre-potentials $H(s, s)$ and $H'(s - 1 , s - 1)$ are likely to be the 
superspin $s$ contributions to a generalization of the conventional $N = 1$, 
$D=4$ conformal supergravity pre-potential. For obvious reasons this 
generalization may be called ``the $N = 1$, $D=4$ conformal superstring 
pre-potential.''   It seems likely that the $N = 1$, $D=4$ conformal 
superstring pre-potential should be the null-string \cite{nul} limit of 
conventional superstring theory.  Finally within an ultimate formulation
of the $N = 1$, $D=4$ superstring, the gauge variations in (\ref{16}) and 
(\ref{21}) should emerge as the different superspin $s$ components of
a ``$N = 1$, $D=4$ conformal superstring group'' that would constitute 
the ``stringy'' extension for the case of and (8) and (9) in the case 
of $s=1$.  The role of the compensating superfields $\G(s-1,s-1)$ or 
$G(s-1,s-1)$ would be to break down this to the ``$N = 1$, $D=4$ 
Poincar\' e superstring group.''

{\bf 3.} Let us turn to the analysis of $N=2$ supersymmetry transformations
in the model described by the action $S^\bot_{s+1/2}+S^\|_s$.  It follows 
from considerations of dimension that such transformations cannot be 
expressed, {\it in manifestly $N=1$ supersymmetric form}, via the linear 
superfields $\G(s-1,s-1)$ and $G'(s,s)$ entering the action functionals, 
but only in terms of their unconstrained potentials $\Phi(s-1,s)$ and 
$\Psi'(s,s-1)$  introduced according to the rules (\ref{6}) and (\ref{7}). 
Accordingly, one then readily arrives at the following supersymmetry 
transformations
\bea
 \d H_\ss &=& -4i\bar\xi_\ad \Psi'_\sr + {\rm c.c.} ~~~, \cr
\d \Phi_\rs &=& \frac{i(s-1)}{2(s+1)} \xi_\a \bar D_\ad D^\b
H'_{\b\a(s-2)\ad(s-1)}
- \frac {is}{2(s+1)} \xi^\b \bar D_\ad D_\b H'_\rr \cr
&+& \frac {is}{s+1} \xi^\a \bar D_\ad \Psi'_\sr
- i \xi^\b D_\b \bar\Psi'_\rs\cr
&+& i\frac{s-1}s \xi_\a D^\b \bar\Psi'_{\b\os} ~~~,  \cr
 \d H'_\rr &=& -4i\bar\xi^\ad \Phi_\rs + {\rm c.c.} ~~~,\cr
\d \Psi'_\sr &=& \frac i2 \xi^\b \bar D^\ad D_\a H_{\b\rs}
- \frac i2 \xi_\b \bar D^\ad D^\b H_\ss\cr
&+& i\frac {s-1}s \xi_\a \bar D^\ad \Phi_\rs
+ i\xi^\b D_\b \bar\Phi_\sr \cr
&+& i\xi^\b D_\a \bar\Phi_{\b\rr} ~~~,
\label{27}
\eea
where $\xi_\a$ is a constant spinor parameter.
These transformations leave the total action $S^\bot_{s+1/2}
+ S^\|_s$ invariant.

We have obtained $N=2$ $(s,s+1/2)$ multiplets described in terms of a
special $N=1$ superfield realization: the longitudinal formulation for 
superspin $s$ and the transversal formulation for superspin $s+1/2$.
It can be seen that there exist three other {\it a priori} combinations 
to describe $N=2$ multiplets from $N=1$ superspin-$s$ and superspin-$(
s+1/2)$ multiplets.  However none seem to allow the construction of an 
$N=2$ multiplet.  As for the possibility of an $N=2$ multiplet constructed 
from $(s-1/2,s)$, all four choices do not lead to non-trivial supersymmetry 
transformations.  In some more detail, the only possibility allowing 
non-trivial mixing of complex potentials via each other, under hypothetical 
supersymmetry transformations, appears to be the same: the longitudinal 
formulation for integer superspin $s$ and the transversal formulation 
for half-integer superspin $s-1/2$.  Here the dynamical superfields are: 
$H(s-1,s-1)$, $\Phi(s-2,s-1)$ (superspin $s-1/2$) and $H'(s-1,s-1)$, 
$\Psi'(s,s-1)$ (superspin $s$). This case differs from that previously 
considered only by superfield tensor structure.  When the highest superspin 
was half-integer, the tensor structures coincided for complex superfields, 
but differed by two indices for the real superfields. When the highest 
superspin is integer, the tensor structures coincide for the real 
superfields and differ by two indices for the complex ones. This fact 
restricts the number of independent candidate structures for $N=2$ 
supersymmetric transformation laws.  Modulo purely gauge variations, 
the general ansatz for such a transformation reads
\bea
\d H_\rr &=& \a_1\xi^\a \Psi'_\sr
+ \a_2\xi^\b D_\a H'_{\b\a(s-2)\ad(s-1)} \cr
&+& \a_3\xi^\b D_\b H'_\rr + {\rm c.c.} ~~~,
\non \\
\d \Phi_{\a(s-2)\ad(s-1)} &=&
i\b_1 \bar\xi_\ad \pa^{\a\bd} H'_{\oo\bd} \cr
&+& i\b_2 \bar\xi_\bd \pa^{\a\bd} H'_\rr
+ \b_3 \bar\xi_\bd \bar D^\bd D^\a H'_\rr \cr
&+& \b_4 \xi^\a D^\a \Psi'_\sr + \b_5 \bar\xi^\ad D^\a \bar\Psi'_\rs ~~~,
\non \\
\d H'_\rr &=& \g_1\xi_\a \Phi_{\a(s-2)\ad(s-1)}
+ \g_2\xi^\b D_\a H_{\b\a(s-2)\ad(s-1)} \cr
&+& \g_3\xi^\b D_\b H_\rr + {\rm c.c.} ~~~,
\non \\
\d \Psi'_\sr &=& i\d_1 \bar\xi_\ad {\pa_\a}^\bd H_{\oo\bd}\cr
&+& i\d_2 \bar\xi_\bd {\pa_\a}^\bd H_\rr
 + \d_3 \bar\xi_\bd \bar D^\bd D_\a H_\rr \cr
&+& \d_4 \xi_\a D_\a \Phi_{\a(s-2)\ad(s-1)}
+ \d_5 \bar\xi_\ad D_\a \bar\Phi_\ro ~~~,
\label{31}
\eea
where $\a_i$, $\b_i$, $\g_i$ and $\d_i$ are arbitrary coefficients.
Surprisingly enough, direct calculations show that the requirement of
invariance of the action $S^\bot_{s-1/2} + S^\|_s$ under this transformation
can be satisfied only when all the coefficients vanish.

To clearly illustrate the difference between the considered $N=2$
supermultiplets, let us investigate the off-shell superspin content 
of the theories (a similar analysis for the multiplet of $N=2$ 
supergravity was given many years ago for the linearized theory 
\cite{gs1,sjg} and later the full non-linear theory \cite{gs2}).
That can be done by decomposing unconstrained and
linear superfields onto irreducible components according to the
rules given in \cite{og}.

We will not reproduce here the decompositions
of the dynamical superfields onto irreducible ones, but only
describe the superspin contents:
\bea
H(s,s) &=& (s+\hf)^r \oplus s \oplus (s-\hf) \oplus \ldots
\oplus \hf \oplus 0 ~~~,
\label{39}          \\
\G(k,l) &=& \frac{k+l+1}2 \oplus \frac{k+l}2 \oplus \ldots
\oplus \frac{k-l}2 \oplus \frac{k-l-1}2~~, \qquad k>l ~~~,
\label{40}\\
\G(k,k) &=& \frac{2k+1}2 \oplus k \oplus \ldots
\oplus \hf \oplus 0 ~~~,
\label{41}\\
G(k,l) &=& \frac{k+l}2 \oplus \frac{k+l-1}2 \oplus \ldots
\oplus \frac{k-l+1}2 \oplus \frac{k-l}2~~, \qquad k\ge l ~~~.
\label{42}
\eea
Here the superscript `$r$' is used for real representations.
It is seen that $G(k,l)$ and $\G(k,l-1)$ describe equivalent
representations.

Now, we are in position to deduce the superspin content in the 
longitudinal and transversal formulations.  For this purpose one should 
eliminate the gauge degrees of freedom of the dynamical superfields. As 
a result, one finds
\bea
S^\bot_{s+1/2} &:& (s + \hf)^r \oplus (s - \hf) \oplus
(s-1) \oplus \ldots \oplus \hf \oplus 0 ~~~,
\label{43}\\
S^\|_{s+1/2} &:& (s + \hf)^r \oplus (s-1) \oplus
(s - \frac32) \oplus \ldots \oplus \hf \oplus 0 ~~~,
\label{44}\\
S^\|_s &:& s \oplus (s - \hf)^r \oplus (s-1) \oplus
(s - \frac32) \oplus \ldots \oplus \hf \oplus 0 ~~~,
\label{45}\\
S^\bot_s &:& (s+\hf) \oplus s \oplus (s-\hf)^r \oplus (s-1) \oplus
(s-\frac32) \oplus \ldots \oplus \hf \oplus 0 ~~~.
\label{46}\eea
It is now obvious that only pairing the formulations $S^\bot_{s+1/2}$ 
and $S^\|_s$ allows the possibility to join the $N=1$ supermultiplets,
entering the actions, into $N=2$ supermultiplets which off-shell
should have the following general structure (see, e.g., \cite{ggrs})
$$
(\l+\hf)^r \oplus \l \oplus (\l-\hf)^r \quad {\rm or} \quad
(\hf)^r \oplus 0 ~~~.
$$

{\bf 4.} Now we turn to the $N=1$, $D=4$ AdS superspace defined by 
the algebra of covariant derivatives $\cD_A = (\cD_a, \cD_\a, \cDB^\ad)$
\bea
& \{{\cD}_\a,\cDB_{\ad}\}=-2i{\cD}_{\a\ad} ~~~, \qquad
[{\cD}_{\a\ad},{\cD}_{\b\bd}]=-2\mb \mu
(\varepsilon_{\a\b} \bar M_{\ad\bd} +
\varepsilon_{\ad\bd} M_{\a\b}) ~~~, \label{47}\\
& \{ {\cD}_\a,{\cD}_\b\}=-4\mb M_{\a\b} ~~~, \qquad
[{\cD}_\a,{\cD}_{\b\bd}]=i\mb
\varepsilon_{\a\b} \cDB_{\bd} ~~~,
\label{48}
\eea
and conjugates to (\ref{48}). Here $M$ and $\bar M$ are the Lorentz 
generators, and $\mu$ the torsion (the square-root of the constant
curvature) of the AdS superspace.  The AdS analogues of the constraints
(\ref{3}) and (\ref{4}) read \cite{is}
\bea
\cDB^\ad \G_\kl = 0
&\Leftrightarrow& (\cDB^2 - 2(l+2)\mu) \G(k,l) = 0 ~~~, \qquad l> 0 ~~~,
\label{49}\\
\cDB_\ad G_\kl = 0
&\Leftrightarrow& (\cDB^2 + 2l\mu) G(k,l) = 0 ~~~,
\label{50}
\eea
and imply some specific features for transversal and longitudinal
superfields.
First, the corresponding subspaces in the space of
tensor superfields of type $(k,l)$ have empty intersection and
supplement each other. This can be proved by checking the
properties of the projectors $\bar \cP_\bot$ and $\bar \cP_\|$ on the
subspaces of transversal and longitudinal linear superfields
respectively
\bea
&{}& \bar \cP_\bot = \frac{\cDB^2 + 2l\mu}{4(l+1)\mu} ~~~,\qquad
\bar \cP_\| = - \frac{\cDB^2 - 2(l+2)\mu}{4(l+1)\mu} ~~~,
\non \\
&{}& \bar \cP_\| + \bar \cP_\bot = 1 ~~~,\qquad \quad
\bar \cP_\| \,  \bar \cP_\bot = \bar \cP_\bot \, \bar \cP_\| = 0 ~~~.
\label{51}
\eea
Second, the equivalence of the spaces of superfields $G(k,l)$ and
$\G(k,l-1)$, we have mentioned after Eq. (\ref{42}),
becomes evident in the AdS superspace from the explicit
form of the intertwining operators:
\be
G_\kl = \cDB_\ad \G_{\a(k)\ad(l-1)} \quad \Leftrightarrow \quad
\G_{\a(k)\ad(l-1)} = -\frac1{2(l+1)\mu} \cDB^\ad G_\kl ~~~.
\label{52}
\ee
The latter relations allow the potentials $\Phi(k,l+1)$ and $\Psi(k,l-1)$,
which are defined in complete analogy with Eqs. (\ref{6}), (\ref{7}), to 
be expressed in terms of their strengths $\G(k,l)$ and $G(k,l)$ modulo 
purely gauge parts.

It was shown in Ref. \cite{ks2} that each of the formulations (\ref{14}),
(\ref{15}), (\ref{19}) and (\ref{20})
has its counterpart in the AdS superspace.
The corresponding actions are obtained from the flat ones
by replacing the flat derivatives $D$'s with the
AdS covariant ones $\cD$'s along with adding $\m$-dependent terms.
Similarly to
the flat superspace, we now consider the $N=2$ supersymmetry
of the action
$S^\bot_{s+1/2}[H,\G] + S_s^\|[H',G']$. For the
second supersymmetry to be covariant with respect to the $N=1$
AdS superalgebra,
it is convenient to pass from constant spinor
parameter $\xi_\a$ to a Killing spinor superfield $\ve_\a$ defined by
\be
\cD_\a \ve_\a = \cDB_\ad \ve_\a = 0 ~~~,\qquad
\mu \cD^\a \ve_\a = \bar \mu \cDB_\ad \vb^\ad ~~~.
\label{52'}
\ee
The description in terms of $\ve_\a$ and its conjugate is equivalent in 
the AdS superspace to the use of a real linear superfield $\ve$ subject
to the constraints
\be
(\cDB^2 -4\m)\ve = \cD_\a \cDB_\ad \ve = 0 \qquad \ve =\bar \ve ~~~.
\label{52''}
\ee
The superfields $\ve_\a$ and $\ve$ are connected as follows
\be
\cD^\a \ve_\a =2\mb \ve \qquad \cD_\a\ve = 2\ve_\a ~~~.
\label{52'''}
\ee
The $N=2$ supersymmetry transformations in the AdS superspace differs 
in form from the flat ones (\ref{27}) only by $\mu$-dependent terms:
\bea
 \d_\ve H_\ss &=& -4i\vb_\ad \Psi'_\sr + {\rm c.c.}~~~,
\non \\
\d_\ve \Phi_\rs &=& \frac{i(s-1)}{2(s+1)} \ve_\a \cDB_\ad \cD^\b
H'_{\b\a(s-2)\ad(s-1)}
- \frac {is}{2(s+1)} \ve^\b \cDB_\ad \cD_\b H'_\rr\cr
&+& 2is \mb \vb_\ad H'_\rr
+ \frac{is}{s+1} \ve^\a \cDB_\ad \Psi'_\sr\cr
&-& i \cD^\b \ve_\b \bar\Psi'_\rs
+ i\frac{s-1}s \ve_\a \cD^\b \bar\Psi'_{\b\os}~~~, \non \\
 \d_\ve H'_\rr &=& -4i\vb^\ad \Phi_\rs + {\rm c.c.}~~~,
 \non \\
\d_ve \Psi'_\sr &=& \frac i2 \ve^\b \cDB^\ad \cD_\a H_{\b\rs}
- \frac i2 \ve_\b \cDB^\ad \cD^\b H_\ss\cr
&+& 2is \mb \vb^\ad H_\ss
+ i\frac{s-1}s \ve_\a \cDB^\ad \Phi_\rs\cr
&+& i\cD^\b \ve_\b \bar\Phi_\sr
+ i\ve^\b \cD_\a \bar\Phi_{\b\rr}~~~.
\label{57}
\eea
The transformations (\ref{57}) leave the action $S^\bot_{s+1/2} + S^\|_s$ 
in the AdS superspace invariant and reduce in the flat limit ($\mu\to 0$) 
to (\ref{27}). It is a property of the AdS superspace that the transformation 
laws can always be expressed in terms of the strengths $\G(s-1,s-1)$ and 
$G'(s,s)$:
\bea
\d _\ve H(s,s) &=& 2i \ve\left(G'(s,s) - \bar G'(s,s)\right)~~~,
\non \\
\d _\ve \G_\rr &=& -\frac i2 (\cDB^2 + 2(s-1)\mu) \ve^\b \cD_\b H'_\rr \cr
&+& 2is\mb \cDB^\ad \vb_\ad H'_\rr
- \frac{2is}{s+1} \ve^\a \cDB^\ad G'_\ss \cr
&-& \frac{is}{s+1} \cDB^\ad \ve \cD^\a \bar G'_\ss~~~,
\non \\
\d _\ve H'(s-1,s-1) &=&
2i \ve\left(\G(s-1,s-1) - \bar \G(s-1,s-1)\right)~~~,
\non \\
\d _\ve G'_\ss &=& -\frac i2 (\cDB^2 - 2(s+2)\mu) \ve^\b \cD_\b H_\ss \non \\
&+& 2is\mb \cDB_\ad \vb^\bd H_{\sr\bd}
- 2i \ve_\a \cDB_\ad \G_\rr \non \\
&-& i \cDB_\ad \ve \cD_\a \bar\G_\rr~~~.
\label{61}
\eea
Here the derivatives are assumed to act on all the objects placed to their 
right. A new feature of the $N=2$ supersymmetry variations is that the 
covariant derivatives enter some expressions in higher powers.  As a 
consequence, we are able to add appropriate variations proportional to 
the equations of motion of the actions $S^\bot_{s+1/2}$ and $S^\bot_s$.
Such a possibility did not arise for the previous expression (\ref{57}) 
of the second supersymmetry nor for the supersymmetry between the 
component actions for arbitrary spin fields \cite{c}.  In fact, to come 
to the latter form of the transformation laws (\ref{61}) one has first 
to express the potentials via their strengths (modulo purely gauge parts), 
in accordance with (\ref{52}), and then to add some purely gauge variations 
as well as certain terms proportional to the equations of motion.

The transformations turn out to form a closed algebra off the mass-shell!
It is an instructive exercise to derive the relation
\be
[ \d_\ve, \d_{\ve'} ] = \d _\cK + \d _g + \d _\g~~~.
\label{62}
\ee
Here $\d_\cK$ denotes an AdS transformation acting on a tensor superfield 
$U$ by the law
\be
\d_\cK U = \cK U ~~~, \qquad
\cK = -\hf k^\aa \cD_\aa + (k^\a \cD_\a
+ k^{\a(2)} M_{\a(2)} + {\rm c.c.})
= \bar\cK ~~~,
\label{63}
\ee
with the parameters constrained by
\bea
k_\a = \frac i8 \cDB^\ad k_\aa~~~, &\qquad& k_{\a(2)} = \cD_\a k_\a~~~,
\non \\
\cDB_\ad k_\aa = \cD_\a k_\aa = 0~~~, &\qquad & \cD^\a \cDB^\ad k_\aa
= \cDB^\ad \cD^\a k_\aa = 0~~~,
\label{64}
\eea
and chosen in the case at hand as follows
\be
k_\aa = 16 i (\ve_\a \vb'_\ad - \ve'_\a \vb_\ad)~~~.
\label{65}
\ee
$\d g$ and $\d \g$ denote special $\ve$-dependent gauge transformations
(\ref{16}--\ref{18}) and (\ref{21}--\ref{23})
\bea
g_\aa &=& (\cDB^2 - 2(s+2)\m)(\ve^\b \ve' - \ve'^{\b} \ve)
\cD_\b H_\ss \non \\
&-& 4s\mb \cDB_\ad (\vb^\bd \ve' - \vb'^\bd \ve) H_{\sr \bd}~~~,
\non \\
\g_\rr &=& (\cDB^2 + 2(s-1)\m)(\ve^\b \ve' - \ve'^{\b} \ve)
\cD_\b H_\rr \non \\
&-& 4s\mb \cDB^\ad (\vb_\ad \ve' - \vb'_\ad \ve) H_{\rr}~~~,
\label{66}
\eea
Eq. (\ref{64}) defines a Killing supervector $\cK$ of the AdS superspace, 
$[\cK, \cD_A]=0$. The set of all Killing supervectors is known to form 
the Grassmann shell of the superalgebra $osp(1,4)$. Remarkably, the union 
of the transformations (\ref{61}) and the $N=1$ AdS ones (\ref{63}) provides 
a realization of the $N=2$ supersymmetry algebra, $osp(4,2)$. The point 
is that the spinor Killing superfield $\ve_\a$ defined by Eq. (\ref{52'})
contains among its components two independent constant parameters: a spinor
and a scalar. The latter proves to correspond to $O(2)$-rotations which 
enter into $osp(4,2)$.

Eqs. (\ref{57}), (\ref{61}) allow us to obtain new representations for $N=2$ 
supersymmetry in flat superspace.  The point is that there are two 
different possibilities for carrying out the limit to flat superspace:
(I) One can keep $\ve$ fixed in the limit $\m \rightarrow 0$; (II)
$\ve $ goes to infinity in the limit $\m \rightarrow 0$ such that
$\bar \m \ve $ remains finite and non-zero. In the former case,
the flat-superspace limit of Eqs. (\ref{52''}), (\ref{52'''}) reads
\be
\ve_\a = \hf D_\a \ve~~~, \qquad \bar D^2 \ve = D_\a \bar D_\ad \ve=0~~~,
\qquad \ve=\bar\ve~~~.
\label{67}
\ee
and $\ve$ has the explicit form
\bea
&{}& \ve= f + 2\q^\a \xi_\a + 2\bar \q_\ad \bar \xi^\ad~~~,  \cr
&{}& f = const~~~, \qquad \xi_\a = const~~~.
\label{68}
\eea
Then (\ref{61}) turns into
\bea
\d _\ve H(s,s) &=& 2i \ve\left(G'(s,s) - \bar G'(s,s)\right)~~~,
\non \\
\d _\ve \G_\rr &=& -\frac i2 \bar D^2  \ve^\b D_\b H'_\rr
 - \frac{2is}{s+1} \ve^\a \bar D^\ad G'_\ss \cr
&-& \frac{is}{s+1} \bar D^\ad \ve D^\a \bar G'_\ss~~~,
\non \\
\d _\ve H'(s-1,s-1) &=& 2i \ve\left(\G(s-1,s-1) - \bar \G(s-1,s-1)\right)~~~,
\non \\
\d _\ve G'_\ss &=& -\frac i2 \bar D^2 \ve^\b D_\b H_\ss
 - 2i \ve_\a \bar D_\ad \G_\rr \cr
&-& i \bar D_\ad \ve D_\a \bar\G_\rr~~~.
\label{69}
\eea
These expressions describe $N=2$ supersymmetry transformations (\ref{27})
but written now in terms of $\G$ and $G$. Due to the explicit dependence
of $\ve$ on $\q$, {\it the transformations obtained do not have a manifestly
$N=1$ supersymmetric form}; however, commuting (\ref{69}) with an $N=1$ 
super-translation results in a purely gauge shift. Similarly, the parameter 
$f$ in (\ref{68}) generates purely gauge transformations.  Now, let us 
consider the second possibility of reduction to the flat superspace. Here 
$\ve_\a$ is well-defined in the limit $\m = \bar \m \rightarrow 0$, 
satisfies the equations
\be
D_\a \ve_\a = \bar D_\ad \ve_\a = 0~~~, \qquad
 D^\a \ve_\a =  \bar D_\ad \vb^\ad~~~.
\label{70}
\ee
and therefore has the following form
\be
\ve_\a = \xi_\a + \theta_\a c~~~, \qquad \xi_\a = const~~~,
\qquad c = const~~~.
\ee
Under this contraction only the transformation law (\ref{57}) possesses
a nonsingular flat-superspace limit which reduces to the replacement
$\cD_A \rightarrow D_A$, $\m \rightarrow 0$.  Choosing $c = 0$ we then 
recover the exact result in Eq. (\ref{27}). For $\xi_\a =0$, on the 
other hand, our transformation law  will describe internal
$O(2)$-rotations of the dynamical superfields in Minkowski superspace!

{\bf 5.} Let us comment on the $N=2$ supersymmetry
between the formulations $S^\|_{s+1/2}[H,G]$ and $S^\bot_s[H',\G']$
dually equivalent to $S^\bot_{s+1/2}[H,\G]$ and $S^\|_s[H',G']$
respectively. It turns out that the use of duality transformations 
does allow us to obtain some symmetry, with a Killing spinor parameter,
of the action $S^\|_{s+1/2} + S^\bot_s$ in the AdS superspace,
but this symmetry will have a nonstandard form. In more detail,
the duality of the actions $S^\bot_{s+1/2}$ and $S^\|_{s+1/2}$
was established in \cite{ks2} with the aid of the auxiliary action
\bea
S_V[H, \G, G, V_\|] &=& S^\bot_{s+1/2}[H,\G] +
\left(-\hf\right)^s \int d^8z E^{-1} \Big\{V_\| \cdot \bar V_\| \cr
&+& (2\bar\G\cdot V_\| + \frac{s+1}s V_\| \cdot V_\| - \frac2s G\cdot V_\|
+ {\rm c.c.})\Big\}~~~.
\label{}\eea
Here $d^8z E^{-1}$ is the invariant measure of the AdS superspace and
both the auxiliary superfield $V_\|$ and dynamical superfield $G$ are
longitudinal linear. Using the equation of motion for $G$, $V_\|$ vanishes 
and the action $S_V$ reduces to $S^\bot_{s+1/2}$. The superfields $\G$ and 
$V_\|$ can be expressed in terms of $G$ and $H$ using their own equations 
of motion, that leads to the formulation $S^\|_{s+1/2}$. Now, we are to 
look for transformation laws of superfields $G$ and $V_\|$ for which the
variation of $S_V$ equals to that of $S^\bot_{s+1/2}$. First we put
$\d V_\| = 0$. Then we find an equation on $\d G$
\bea
\d S_V - \d S^\bot_{s+1/2} = \left(-\hf\right)^s \int d^8z E^{-1}
(2\d \bar\G - \frac2s\d G)\cdot V_\| = 0~~~.
\label{}\eea
If the quantity in the parentheses is transversal linear, the integrand 
will be a total derivative. This uniquely requires
\bea
\d G = s \bar \cP_\| \,\d \bar\G~~~,
\label{}\eea
where the projector $\bar \cP_\|$ was defined in Eq. (\ref{51}).  The 
variation $\d\G'$ of the formulation $S^\bot_s[H,\G']$ can be determined 
analogously, and we obtain a symmetry between the auxiliary actions 
for superspins $s+1/2$ and $s$. However, after eliminating $\G$,
$G'$ and the auxiliary superfields, with the use of their equations
of motion, we will obtain rather subtle expressions for the transformation 
laws of the rest superfields which involve higher derivatives, 
$1/\mu$-factors and do not admit flat-superspace limit. 

Finally, let us end by noting that the investigation of the existence 
of these massless arbitrary superspin-$s$ $N = 2$, $D=4$ multiplets 
is not an idle concern from the view of the $N = 1$, $D=4$ superstring. 
It has long been known that massive representations of $N = 1$, $D=4$ 
supersymmetry are equivalent to $N = 2$ representations.  Thus, our 
confirmation that the $N=2$ representations exist for arbitrary 
superspin-$s$ gives a positive indication for the existence of 
{\it {massive}} $N = 1$, $D=4$ arbitrary superspin-$s$ multiplets.  
Since all $N = 1$, $D=4$, $s >1$ supermultiplets must be massive in 
the case of the superstring, our present work is seen to be 
consistent with our conjecture that the pre-potentials $H(s,s)$ 
and $H'(s-1,s-1)$ can be consistently interpreted as different
components of a superstring theory.

\vspace{1cm}

\noindent
{\bf Acknowledgements:}  This work was supported in part by the Russian
Foundation for Basic Research Grant No. 96-02-16017. The research of SMK
was supported in part by the Alexander von Humboldt Foundation. The work
of AGS was also supported by a Fellowship of Tomalla Foundation (under
the program of International center for Fundamental Physics in Moscow).
The research of SJG was supported by the U.S. NSF under grant 
PHY-96-43219 and NATO under grant CRG-93-0789.

\end{document}